# Analysis of Raman modes in Mn-doped ZnO nanocrystals

**Shuxia Guo**[*,1,2], **Zuliang Du**[2], **and Shuxi Dai**[2]

[1] Department of Physics, Jiaozuo Teacher's College, Jiaozuo 454001, P. R. China
[2] Key Laboratory for Special Functional Materials of Ministry of Education, Henan University, Kaifeng 475004, P. R. China

Mn-doped ZnO was synthesized using a co-precipitation technique. X-ray diffraction (XRD) measurements and photoluminescence (PL) spectra show that Mn ions are doped into the lattice positions of ZnO. The modes at 202, 330, and 437 cm$^{-1}$ in the Raman spectrum are assigned as 2E$_2$ (low), E$_2$ (high)–E$_2$ (low), and E$_2$ (high) modes of ZnO base, respectively. The mode at 528 cm$^{-1}$ is ascribed to a local vibrational mode related to Mn. The mode at 580 cm$^{-1}$ should be an intrinsic mode of ZnO and assigned to E$_1$ longitudinal optical (LO). Its reinforcement should result from a combination of resonance at the excitation wavelength and impurity-induced scattering.

## 1 Introduction

ZnO offers some unique optoelectronic properties due to its wide bandgap of 3.4 eV and large excitonic binding energy of 60 meV. The material was predicted to show room temperature ferromagnetism when doped with 5% Mn [1]. Numerous studies of Mn-doped ZnO have presented diverse and controversial results. It is difficult to reach consensus as to whether ferromagnetism arises from the intrinsic properties of the compound or extrinsic ferromagnetic impurities [2, 3]. Thus, investigation of its microstructure is imperative. Raman scattering can yield important information about the nature of a solid on a scale of the order of a few lattice constants, so it can be used to study the microscopic nature of structural and/or topologic disorder.

The Raman mode at about 580 cm$^{-1}$ has been observed by many groups who studied the phonon properties of ZnO-based materials, such as ZnO nanostructures and intentionally doped ZnO samples. Various assignments and activation mechanisms for the mode have been put forward as follows: (i) a wurtzite–ZnO silent mode allowed due to the breakdown of the translational crystal symmetry induced by defects and impurities [4]; (ii) a longitudinal optical (LO) phonon mode due to spatial confinement within the dot boundaries [5]; (iii) a mode induced by defects, such as oxygen vacancies [6]; and (iv) a quasimode of mixed A$_1$ and E$_1$ symmetry due to crystallite-orientation effects [7]. The mode falls in the region of LO phonons which, due to their long-range electrostatic field, can couple with electrons and modify the optoelectronic characteristics of ZnO-based materials. So investigation of the mode is of great importance.

In the work reported here, we fabricated pure ZnO and Mn-doped ZnO via an oxalate co-precipitation technique. X-ray diffraction (XRD) shows that no second phase appears in the Mn-doped ZnO. Also, room temperature photoluminescence (PL) suggests that Mn substitutionally replaces Zn in ZnO. The mode at 580 cm$^{-1}$ was observed in the Raman spectrum of Mn-doped ZnO. It is more reasonable to ascribe the mode to an LO phonon mode with $E_1$ symmetry, as will be discussed subsequently.

**2 Experimental** Polycrystalline Zn$_{1-x}$Mn$_x$O powders were prepared using a co-precipitation technique. Oxalate precursors Zn$_{1-x}$Mn$_x$(C$_2$O$_4$) · 2H$_2$O were obtained by adding 100 ml aqueous solution of oxalic acid (0.07 mol) into 100 ml aqueous solutions of zinc and manganese acetates (0.07 mol cations). The precipitates were collected, washed with copious quantities of deionized water, and dried in air at 373 K. Zn$_{1-x}$Mn$_x$O powders were obtained by thermal decomposition of the white precursors at 673 K for 3 h in air.

All samples were characterized using powder XRD with an X'Pert Philips diffractometer and Ni-filtered Cu Kα radiation, operating at 40 kV and 40 mA and using a 0.01° scan step. The mean crystallite size was estimated through the Scherrer equation. Absorption spectra were measured with a Cary 5000 spectrometer. PL spectra were acquired with a SPEX F212 fluorescence spectrophotometer at room temperature. Raman spectra were obtained at room temperature under a backscattering geometry using a Renishaw RM-1000. The 632.8 nm line of an He–Ne laser was used for excitation.

**3 Results and discussion** The XRD patterns of ZnO and $Zn_{0.99}Mn_{0.01}O$ powders prepared using the co-precipitation technique are shown in Fig. 1. All diffraction peaks from the samples correspond to the hexagonal ZnO structure. Compared with pure ZnO (33.2 nm), the crystalline particle size of Mn-doped ZnO is smaller, i.e., 17.9 nm; however, its lattice parameters are larger. And the peaks exhibit significant broadening in the $Zn_{0.99}Mn_{0.01}O$ XRD pattern. Such large XRD line broadening in transition-metal-doped ZnO indicates lattice defect formation in the alloys.

Photoluminescence measurements were performed to confirm the Mn position in the host lattice. Figure 2 shows the PL spectra of Mn-doped ZnO and pure ZnO excited at 330 nm at room temperature. The ZnO spectrum shows two emission bands: an exciton emission band (weak and narrow) and a trap emission band (strong and broad). The fluorescence band for Mn-doped ZnO consists of four peaks extracted by Gauss fitting. Compared with the PL spectrum of pure ZnO, the two peaks at about 385 and 520 nm should result from the bandgap and defect trap emissions, respectively. The other two peaks should be attributed to the $^4E(G) \rightarrow {}^6A_1(S)$ and $^4T_2(G) \rightarrow {}^6A_1(S)$ transitions of $Mn^{2+}$ ions [8]. This behavior indicates that the Mn atoms have been doped into some lattice sites of ZnO [9].

The space group of wurtzite ZnO belongs to $C_{6v}^4$ (P63mc), with two formula units per primitive cell and with all atoms occupying $C_{3v}$. ZnO has phonon-dispersion relations consisting of 12 branches whose group-theoretical analysis at the Brillouin zone center ($q = 0$) yields a decomposition into the following phonon modes: $\Gamma = 2 \times (A_1 + B_1 + E_1 + E_2)$. Among these modes, there are acoustic modes with $\Gamma_{aco} = A_1 + E_1$ and optical modes with $\Gamma_{opt} = A_1 + (2 \times B_1) + E_1 + (2 \times E_2)$. Of these, the $B_1$ modes are silent modes, and other modes are Raman active. The phonons with $A_1$ and $E_1$ symmetry are polar and hence exhibit different frequencies for the transverse optical (TO) and LO phonons due to the long-range electrostatic forces. The $A_1$ phonon vibration is polarized parallel to the C-axis; the $E_1$ phonon is polarized perpendicular to the C-axis. Every mode corresponds to a band in the Raman spectrum. The intensities of these bands depend on the scattering cross-section of these modes.

The Raman spectra of Mn-doped ZnO and ZnO nanoparticles in the range 280–850 cm$^{-1}$ are shown in Fig. 3. The spectrum in the top panel is that of Mn-doped ZnO, and in the bottom panel is that of ZnO. Both Raman spectra are dominated by three peaks at about 202, 330, and 437 cm$^{-1}$. The mode at 437 cm$^{-1}$ is assigned to $E_2$ (high) mode, and the modes at 202 and 330 cm$^{-1}$ are assigned as $2E_2$ (low) and $E_2$ (high)–$E_2$ (low) modes, respectively [10]. From Fig. 3, it is seen that although most of the Raman peaks of Mn-doped ZnO correspond well to those of ZnO, the Raman spectrum of Mn-doped ZnO shows its own characteristics. One of the features is that the intensity of the 437 cm$^{-1}$ mode decreases owing to the potential fluctuations of the alloy disorder. Another feature is the Raman band at about 500–700 cm$^{-1}$ becomes stronger. The band is composed of three peaks, 528, 580, and 662 cm$^{-1}$ extracted by Lorentz fitting (Fig. 3, top panel). The Raman signal from pure ZnO is enlarged in the region from 500 to 700 cm$^{-1}$ in the inset of Fig. 3, and peaks at about 560 and 660 cm$^{-1}$ can be observed.

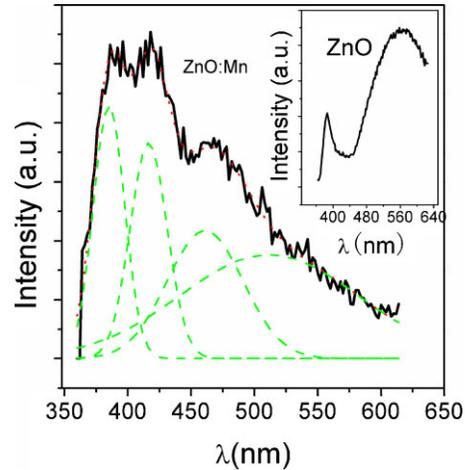

**Figure 2** (online color at: www.pss-b.com) PL spectrum of $Zn_{0.99}Mn_{0.01}O$ excited at 330 nm at room temperature. Data curve is black; the dotted curve is a best fit to the data; and the sum of the individual peaks is shown as dashed curves. Inset: PL spectrum of ZnO.

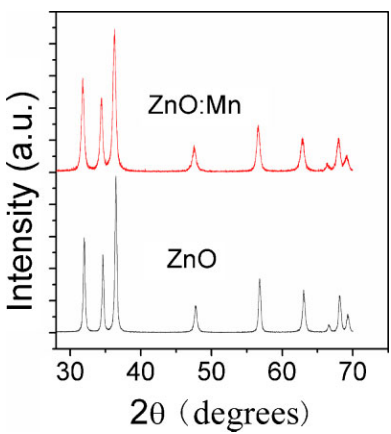

**Figure 1** (online color at: www.pss-b.com) XRD patterns for ZnO and $Zn_{0.99}Mn_{0.01}O$ powders.

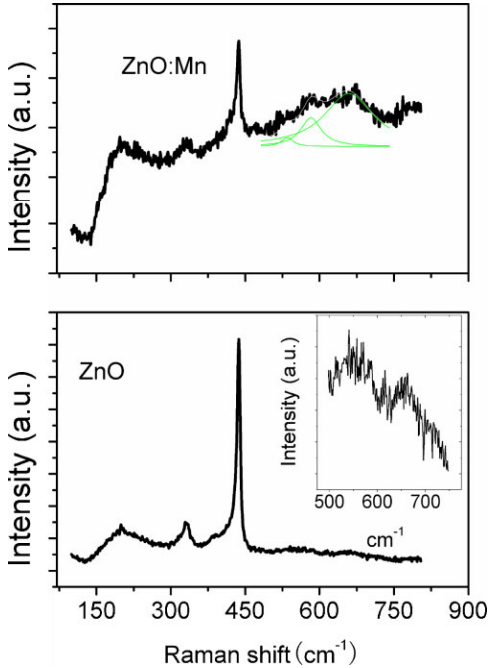

**Figure 3** (online color at: www.pss-b.com) Room temperature Raman spectra of $Zn_{0.99}Mn_{0.01}O$ (top) and ZnO (bottom) powders.

The additional vibrational mode at 528 cm$^{-1}$ should be ascribed to a local vibrational mode related to Mn substituting Zn in ZnO base because it is absent in the Raman spectra of the pure ZnO- and Mn-based compounds [11].

The mode at 580 cm$^{-1}$ was also observed in the spectra of Fe-, Sb-, and Al-doped ZnO thin films [12] and in the spectra of N-doped samples [13]. So it is not a local vibrational mode related to Mn. It is also impossible to believe the mode is induced by oxygen vacancies [6] because the PL band induced by oxygen vacancies appears clearly in the spectrum of pure ZnO, but the mode is not observed in ZnO Raman spectrum.

The mode at 580 cm$^{-1}$ falls between the $A_1$ (LO) at 568 cm$^{-1}$ and the $E_1$ (LO) at 586 cm$^{-1}$ of the bulk ZnO. Theoretical work and experimental results on phonon confinement in ZnO quantum dots indicated that the principal confined LO mode frequency should be between those of the $A_1$ (LO) and the $E_1$ (LO) of the bulk material [5, 14]. Although Mn does induce a decease of the crystalline particle size, as mentioned above, the crystallites studied here are still much larger than those of quantum dots studied in the literature, which are about 4 nm [5]. From Fig. 2, the near band edge emission energies of Mn-doped ZnO and pure ZnO are similar, both at about 3.2 eV, indicating a lack of confinement in the electronic state of the crystallites. Thus, we preclude any significant confinement in the phonon states. On the other hand, a mixed mode or quasi-LO mode may also occur around 580 cm$^{-1}$ near the $A_1$ and $E_1$ LO modes of ZnO [7]. The random orientation nature of a nanocrystallite ensemble, which changes with preparation method, defects and impurities, not only leads to the emergence of mixed-symmetry modes, but also determines the positions of the mixed-symmetry modes. However, Fig. 1 shows no visible changes in the number and relative intensities of peaks following Mn incorporation in ZnO, except for a broadening of the diffraction features. Therefore, the crystallite orientation may not be responsible for the mode at 580 cm$^{-1}$.

The mode at 580 cm$^{-1}$ in the Raman spectrum of Mn-doped ZnO should be ascribed to $E_1$ (LO), which is enhanced and shifted to lower wavenumber due to a resonance at the excitation wavelength and the presence of impurity-induced scattering. The $E_1$ (LO) mode shows resonance Raman enhancement in ZnO when the laser frequency approaches the bandgap $E_0$ [7]. Although the frequency of the laser used here is quite far from the $E_0$ edge of ZnO, there are several reasons that support our assignment above. First, investigation of the $E_1$ (LO) mode of ZnO at room temperature indicated that while 1.92–2.57 eV was quite far from the $E_0$ edge of ZnO, resonance effects could be identified in it [15]. Second, the absorption spectrum of Mn-doped ZnO (Fig. 4) shows that there exists quite weak absorption at 633 nm (i.e., excitation wavelength). The origin of the absorption cannot be deduced from the present data, but its existence ensures LO mode resonance is possible because fundamental resonance effects in light scattering occur for photons with frequencies of the electronic absorption bands. Third, a strong resonance enhancement of the LO mode near 580 cm$^{-1}$ has been observed even for the 568 nm laser line for Mn-doped ZnO [16]. Finally, the resonance enhancement at longer wavelengths without the benefit from well-known exciton effects has been observed in other materials [17].

The contribution from the impurity-induced mechanism is significant for one-LO-phonon scattering, but it can be neglected for scattering by two and more LO phonons [18], which is only due to the intrinsic intraband Fröhlich mechanism. The impurity-induced enhancement for one-LO-phonon scattering results for the following reasons. Serious breakdown of the wave vector conservation because of scattering by random impurities allows phonons with larger

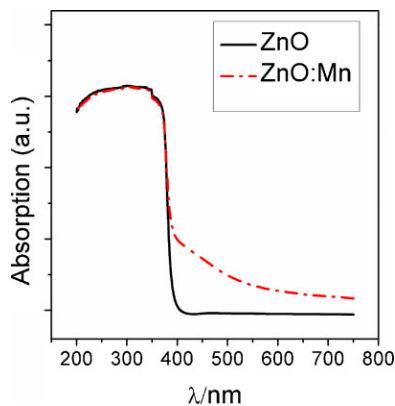

**Figure 4** (online color at: www.pss-b.com) Absorption spectra of ZnO and $Zn_{0.99}Mn_{0.01}O$.

wave vectors to participate in the Raman scattering. These larger wave vectors greatly enhance the intraband Fröhlich contribution, and therefore increase the corresponding scattering cross-section.

The vibrational mode at $660\,cm^{-1}$ is an intrinsic mode of ZnO and is assigned to TA + LO [19]. The mode is also enhanced in the spectrum of Mn-doped ZnO, which should be related to reinforcement of LO phonon scattering. It implies that the mode at $580\,cm^{-1}$ as a silent mode $B_1$ (high) is impossible.

The mode at about $560\,cm^{-1}$ in the Raman spectrum of ZnO may be attributed to $B_1$ (high) based on recent inelastic neutron scattering measurements [20] and *ab initio* calculations [19].

**4 Conclusions** X-ray diffraction and PL spectra indicate the formation of polycrystalline Mn-substituted ZnO without a second phase using the co-precipitation technique. The mode at $580\,cm^{-1}$ in the Raman spectrum of Mn-doped ZnO should be an intrinsic mode of ZnO and assigned to $E_1$ (LO). Its peak shifts towards lower wavenumber due to impurity effects. Its reinforcement should result from a combination of resonance at excitation wavelength and impurity-induced scattering.

**Acknowledgements** This work was supported by the Natural Science Foundation of China (no. 90306010), Program for New Century Excellent Talents in University (no. NCET-04-0653) and State Key Basic Research "973" Plan of China (no. 2007CB616911).